\documentstyle[twocolumn,aps,graphicx]{revtex}

\draft

\begin{document}
\title{Renormalization group flow with unstable particles}
\author{O.A.~Castro-Alvaredo$^\sharp$ and A.~Fring$\,^\star$}
\address{$^\sharp$Departamento de F\'\i sica de Part\'\i culas, Universidad
de Santiago de Compostela, E-15706 Santiago de Compostela, Spain\\
$^\star$Institut f\"ur Theoretische Physik, Freie Universit\"at
Berlin, Arnimallee 14, D-14195 Berlin, Germany }
\date{\today}
\maketitle

\begin{abstract}
The renormalization group flow of an integrable two dimensional quantum
field theory which contains unstable particles is investigated. The analysis
is carried out for the Virasoro central charge and the conformal dimensions
as a function of the renormalization group flow parameter. This allows to
identify the corresponding conformal field theories together with their
operator content when the unstable particles vanish from the particle
spectrum. The specific model considered is the $SU(3)_{2}$-homogeneous
Sine-Gordon model.
\end{abstract}

\pacs{PACS numbers: 11.10Hi, 11.10Kk, 11.30Er, 05.70.Jk}


The study of two-dimensional quantum field theories (2D-QFT) has turned out
to be a fruitful venture since almost three decades. In particular when
exploiting integrability many non-perturbative methods have been developed
over the years. Besides the challenge to understand the underlying
mathematical structures and the intriguing physical applications in two
dimensions itself, e.g. to describe measurable quantities of carbon
nanotubes \cite{Nano}, the ultimate goal is to extrapolate ones findings to
higher dimensions. In particular for the celebrated c-theorem of
Zamolodchikov \cite{ZamC}, which originally describes the renormalization
group trajectory of a function which at the renormalization group fixed
point corresponds to the Virasoro central charge, various counterparts have
been developed in higher dimensions, e.g. \cite{C4D}.

Fairly recently a class of massive integrable quantum field theories, the
homogeneous Sine-Gordon models (HSG) \cite{HSG}, has been proposed
introducing the feature of possessing unstable particles inside its particle
spectrum. Despite the fact that theories containing resonances have been
treated before in the context of two-dimensional massive quantum field
theories, e.g. \cite{roaming}, the HSG-models are somewhat special since
they constitute the first examples of theories which admit a well-defined
Lagrangian description. In general the HSG models are associated to
integrable perturbations of $G$-parafermions of level $k$ \cite{Gep}, i.e.
WZNW-coset theories of the form $G_{k}/U(1)^{\ell }$ with $\ell $ being the
rank of a compact Lie group $G$. As free parameters the model contains $\ell 
$ different mass scales and $\ell -1$ different scales for the resonance
parameter $\sigma $ which enter the Breit-Wigner formula \cite{BW}. In
general an unstable particle of type $\tilde{c}$ is described by
complexifying the physical mass of a stable particle by adding a decay width 
$\Gamma _{\tilde{c}}$, such that it corresponds to a pole in the S-matrix as
a function Mandelstam $s$ at $s=M_{R}^{2}=(M_{\tilde{c}}{}-i\Gamma _{\tilde{c%
}}/2)^{2}$ (for a more detailed discussion see e.g. \cite{ELOP}). As
mentioned in \cite{ELOP} whenever $M_{\tilde{c}}\gg \Gamma _{\tilde{c}}$,
the quantity $M_{\tilde{c}}$ admits a clear cut interpretation as the
physical mass. However, since this assumption is only required for
interpretational reasons we will not rely on it. Transforming as usual in
this context from $s$ to the rapidity plane and describing the scattering of
two stable particles of type $a\,$and $b$ with masses $m_{a}$ and $m_{b}$ by
an S-matrix $S_{ab}(\theta )$ as function of the rapidity $\theta $, the
resonance pole is situated at $\theta _{R}=\sigma -i\bar{\sigma}$.
Identifying the real and imaginary parts of the pole then yields 
\begin{eqnarray}
M_{\tilde{c}}^{2}{}-\frac{\Gamma _{\tilde{c}}^{2}{}}{4}
&=&m_{a}^{2}{}+m_{b}^{2}{}+2m_{a}m_{b}\cosh \sigma \cos \bar{\sigma}
\label{BW1} \\
M_{\tilde{c}}\Gamma _{\tilde{c}} &=&2m_{a}m_{b}\sinh |\sigma |\sin \bar{%
\sigma}\,\,.  \label{BW2}
\end{eqnarray}
Eliminating the decay width from (\ref{BW1}) and (\ref{BW2}), we can express
the mass of the unstable particles $M_{\tilde{c}}$ in the model as a
function of the masses of the stable particles $m_{a},m_{b}$ and the
resonance parameter $\sigma $. Assuming $\sigma $ to be large this gives 
\begin{equation}
M_{\tilde{c}}^{2}\sim \frac{1}{2}m_{a}m_{b}(1+\cos \bar{\sigma}%
)\,\,e^{|\sigma |}\,.  \label{Munst}
\end{equation}

One recognizes the occurrence of the variable $me^{|\sigma |/2}$, which was
introduced originally in \cite{triZam} in order to describe massless
particles, i.e. one may perform safely the limit $m\rightarrow 0,\sigma
\rightarrow \infty $ and one might therefore be tempted to describe flows
related to (\ref{Munst}) as massless flows. In \cite{CFKM} the relative mass
scales between the unstable and stable particles and the stable particles
themselves were investigated by computing the finite size scaling function
from the thermodynamic Bethe ansatz (TBA). A consistent physical picture was
obtained for the overall identification of the flow between different coset
models. It remained, however, an open question how to identify the operator
content. In general this question is left unanswered in the context of the
TBA. For theories with certain properties, it is sometimes possible to
determine at least the dimension of the perturbing operators by
investigating periodicities in the so-called Y-systems \cite{ZAMY}.
Resorting to a different method, namely by appealing to sum rules which are
expressible in terms of correlation functions, the major part of the
operator content was successfully identified for some of the HSG models \cite
{CF}. The purpose of this manuscript is on one hand to confirm and refine
the TBA results by the latter method, i.e. by investigating the
renormalization group flow described by the Zamolodchikov c-function \cite
{ZamC}. We will precisely study the onset of the mass scale of the unstable
particles and investigate how a particular coset flows to another one. On
the other hand, we study in addition the flow of the operator content of one
conformal field theory to another one by exploiting the flow provided by the 
$\Delta $-sum rule of Delfino, Simonetti and Cardy \cite{DSC}.

Denoting by $r$ the radial distance and by $t=\ln r^{2}$ the renormalization
group parameter, the functions $c(t)$ and $\Delta (t)$ were defined in \cite
{ZamC} and \cite{DSC}, respectively, obeying the differential equations 
\begin{eqnarray}
\frac{dc(t)}{dt} &=&-\frac{3}{4}\,e^{2t}\left\langle \Theta (t)\Theta
(0)\right\rangle  \label{dc} \\
\frac{d\Delta (t)}{dt} &=&\frac{1}{\left\langle {\cal O}(0)\right\rangle }%
e^{t}\,\left\langle \Theta (t){\cal O}(0)\right\rangle \,\,.  \label{dd}
\end{eqnarray}
The r.h.s. of these equations involve the two-point correlation functions of
the trace of the energy-momentum tensor $\Theta $ and an operator ${\cal O}$%
, which is a primary field in the sense of \cite{BPZ}. In general these
equations are integrated from $t=-\infty $ to $t=\infty $ and one
consequently compares the difference between the ultraviolet and the
infrared fixed points. In order to exhibit the quantitative onset of the
mass scale of the unstable particles we integrate these equations instead
from some finite value $t_{0}$ to infinity. Restricting our attention to
purely massive theories we use the fact that for those theories the infrared
central charges are zero, such that 
\begin{equation}
c(r_{0})=\frac{3}{2}\int\limits_{r_{0}}^{\infty }dr\,r^{3}\,\,\left\langle
\Theta (r)\Theta (0)\right\rangle \,\,.  \label{cr}
\end{equation}
Instead of the integral representation (\ref{cr}), the $c$-function is
equivalently expressible in terms of a sum of correlators involving also
other components of the energy momentum tensor \cite{ZamC}. In deriving (\ref
{dc}) these terms have been eliminate by means of the conservation law of
the energy momentum tensor. We find (\ref{cr}) most convenient. The flow of $%
c(r_{0})$ will surpass various steps: Starting with $r_{0}=0$ the theory
will leave its ultraviolet fixed point and at a certain definite value, say $%
r_{0}=r_{u}$, the unstable particle will become massive such that $%
c(r_{0}>r_{u})$ can be associated to a different conformal field theory. It
appears natural to identify the mass $M_{\tilde{c}}$ as the point at which $%
c(r_{0})$ is half the difference between the two coset values of $c$. As a
consequence of (\ref{Munst}) we may relate the masses of the unstable
particles at different values of the resonance parameter $\sigma $, $\sigma
^{\prime }$ and expect $M_{\tilde{c}}(r_{u},\sigma )=M_{\tilde{c}%
}(r_{u}^{\prime },\sigma ^{\prime })$. We will employ the latter equality
evaluated in the form (\ref{Munst}) not only as a consistency requirement,
but also as a confirmation of the fact that the renormalization group flow
is indeed achieved by $m\rightarrow r_{0}\,m$. Increasing $r_{0}$ further,
the energy scale of the stable particles will eventually be reached at, say
at $r_{0}=r_{a},r_{b},\ldots ,r_{n}$. Depending on the relative mass scales
between the stable particles these points may coincide. Finally the flow
will reach its infrared fixed point $c(r_{0}=r_{ir})=0$.

Likewise we can integrate equation (\ref{dd}) 
\begin{equation}
\Delta (r_{0})=-\frac{1}{2\left\langle {\cal O}(0)\right\rangle }%
\int\limits_{r_{0}}^{\infty }dr\,r\,\,\left\langle \Theta (r){\cal O}%
(0)\right\rangle \,\,,  \label{dr}
\end{equation}
which allows to keep track of the manner the operator contents of the
various conformal field theories are mapped into each other. We used that
all conformal dimensions vanish in the infrared limit. Fortunately, we have $%
\left\langle \Theta (r){\cal O}(0)\right\rangle \sim $ $\left\langle {\cal O}%
(0)\right\rangle $ in many applications such that the vacuum expectation
value $\left\langle {\cal O}(0)\right\rangle $ cancels often. One should
note, however, that (\ref{dr}) is only applicable to those operators for
which its two-point correlator with the trace of the energy momentum tensor
is non-vanishing, such that one may not be in the position to investigate
the flow of the entire operator content by means of (\ref{dr}).

In order to evaluate (\ref{cr}) and (\ref{dr}) we have to compute the
two-point correlation functions in some way. In 2D-QFT this is probably most
efficiently achieved, by expanding them in terms of n-particle form factors,
i.e. the matrix elements of some local operator ${\cal O}(\vec{x})$ located
at the origin between a multiparticle in-state and the vacuum denoted by $%
\left\langle 0|{\cal O}(0)|V_{\mu _{1}}(\theta _{1})V_{\mu _{2}}(\theta
_{2})\ldots V_{\mu _{n}}(\theta _{n})\right\rangle _{\text{in}}=:F_{n}^{%
{\cal O}|\mu _{1}\ldots \mu _{n}}(\theta _{1},\ldots ,\theta _{n})$. Here
the $V_{\mu }(\theta )$ are some vertex operators representing a particle of
species $\mu $. Abbreviating the sum of the on-shell energies as $%
E=\sum\nolimits_{i=1}^{n}m_{\mu _{i}}\cosh \theta _{i}$, one may write 
\begin{eqnarray}
&&\left\langle {\cal O}(r){\cal O}^{\prime }(0)\right\rangle
=\sum_{n=1}^{\infty }\sum_{\mu _{1}\ldots \mu _{n}}\int\limits_{-\infty
}^{\infty }\frac{d\theta _{1}\ldots d\theta _{n}}{n!(2\pi )^{n}}e^{-r\,E}
\label{corre} \\
&&\times \,\,F_{n}^{{\cal O}|\mu _{1}\ldots \mu _{n}}(\theta _{1},\ldots
,\theta _{n})\,\left( F_{n}^{{\cal O}^{\prime }|\mu _{1}\ldots \mu
_{n}}(\theta _{1},\ldots ,\theta _{n})\,\right) ^{\ast }.  \nonumber
\end{eqnarray}
Using this expansion we replace the correlation functions in the expression
of the c-function $c(r_{0})$ and the scaled conformal dimension $\Delta
(r_{0})$ and perform the r integrations thereafter. Thus we obtain 
\begin{eqnarray}
&&c(r_{0})=3\sum_{n=1}^{\infty }\sum_{\mu _{1}\ldots \mu
_{n}}\int\limits_{-\infty }^{\infty }\frac{d\theta _{1}\ldots d\theta _{n}}{%
n!(2\pi )^{n}}e^{-r_{0}\,E}  \label{cr0} \\
&&\times \left| F_{n}^{\Theta |\mu _{1}\ldots \mu _{n}}(\theta _{1},\ldots
,\theta _{n})\right| ^{2}\frac{(6+6r_{0}E+3r_{0}^{2}E^{2}+r_{0}^{3}E^{3})}{%
2E^{4}}  \nonumber
\end{eqnarray}
and 
\begin{eqnarray}
&&\Delta (r_{0})=-\sum_{n=1}^{\infty }\sum_{\mu _{1}\ldots \mu
_{n}}\int\limits_{-\infty }^{\infty }\frac{d\theta _{1}\ldots d\theta _{n}}{%
n!(2\pi )^{n}}\frac{(1+r_{0}E)e^{-r_{0}\,E}}{2E^{2}}  \label{dr0} \\
&&\,\,\,\times F_{n}^{\Theta |\mu _{1}\ldots \mu _{n}}(\theta _{1},\ldots
,\theta _{n})\left( F_{n}^{{\cal O}|\mu _{1}\ldots \mu _{n}}(\theta
_{1},\ldots ,\theta _{n})\,\right) ^{\ast }\,.  \nonumber
\end{eqnarray}
We will now analyze (\ref{cr0}), (\ref{dr0}) and (\ref{Munst}) for the $%
SU(3)_{2}$-HSG model. This model contains only two self-conjugate solitons
which we denote by ``$+$'', ``$-$'' and one unstable particle, which call $%
\tilde{u}$. The corresponding scattering matrix was found \cite{HSGS} to be $%
S_{\pm \pm }=-1$, $S_{\pm \mp }(\theta )=\pm \tanh \left( \theta \pm \sigma
-i\pi /2\right) /2$, which means the resonance pole is situated at $\theta
_{R}=\mp \sigma -i\pi /2$. Stable bound states may not be formed. Note that
for the corresponding value of  $\bar{\sigma}=\pi /2$ and arbitrary $\sigma $
the condition $M_{\tilde{u}}\gg \Gamma _{\tilde{u}}$ is not fulfilled.
However, as indicated above this condition only helps for a clearer
identification of the mass parameter. For the HSG-models this condition
starts to hold when the level is large, which indicates that in these type
of models this interpretation is in fact a semi-classical one.

A huge class of form factors corresponding to various operators related to
this model were constructed in \cite{CFK,CF}. Labelling an operator by four
quantum numbers $\mu ,\nu ,\tau ,\tau ^{\prime }$ the general n-particle
solution reads 
\begin{eqnarray}
&&F_{2s+\tau ,2t+\tau ^{\prime }}^{{\cal O}_{\tau ,\tau ^{\prime }}^{\mu
,\nu }|M^{+}M^{-}}(\theta _{1},\ldots ,\theta _{n})=H_{2s+\tau ,2t+\tau
^{\prime }}^{{\cal O}_{\tau ,\tau ^{\prime }}^{\mu ,\nu
}|M^{+}M^{-}}\,\!\!\det {\cal A}_{2s+\tau ,2t+\tau ^{\prime }}^{\mu ,\nu } 
\nonumber \\
&&\left( \sigma _{2s+\tau }^{+}\right) ^{s-t+\frac{\tau -1-\nu }{2}}\left(
\sigma _{2t+\tau ^{\prime }}^{-}\right) ^{\frac{1+\tau -\tau ^{\prime }-\mu 
}{2}-t}\prod_{i<j}\hat{F}^{\mu _{i}\mu _{j}}(\theta _{ij})\,.
\end{eqnarray}
We used here a particular ordering by starting with $2s+\tau $ particles of
the type $\mu =+$ followed by $2s+\tau ^{\prime }$ particles of the type $%
\mu =-$, collected in the sets $M^{\pm }=\{\pm ,\ldots ,\pm \}$. Once these
expressions are known, all other form factors related to it by permutations
of the particles may be constructed trivially by exploiting Watson's
equations \cite{Kar}, see \cite{CFK,CF} for details concerning the
HSG-models. The functions $\hat{F}^{\mu _{i}\mu _{j}}$ for all combinations
of the $\mu $'s are 
\begin{eqnarray}
\hat{F}^{\pm \pm }(\theta ) &=&-i/2\tanh \frac{\theta }{2}\exp (\mp \theta
/2) \\
\hat{F}^{\pm \mp }(\theta ) &=&2^{\frac{1}{4}}e^{%
{\textstyle{i\pi (1\mp 1)\pm \theta  \over 4}}%
-%
{\textstyle{G \over \pi }}%
-\int\nolimits_{0}^{\infty }%
{\textstyle{dt \over t}}%
{\textstyle{\sin ^{2}\left( (i\pi -\theta \mp \sigma )\frac{t}{2\pi }\right)  \over \sinh t\cosh t/2}}%
},  \label{14}
\end{eqnarray}

\noindent with $G=0.91597\ldots $ being the Catalan constant. The ($t+s$)$%
\times $($t+s$)-matrix 
\begin{equation}
\left( {\cal A}_{2s+\tau ,2t+\tau ^{\prime }}^{\mu ,\nu }\right) _{ij}=%
{\sigma _{2(j-i)+\mu }^{+}\text{,\quad }1\leq i\leq t \atopwithdelims\{. \hat{\sigma}_{2(j-i)+2t+\nu }^{-}\,\,\,\text{,\quad }t<i\leq s+t}%
\label{Acomp}
\end{equation}
has as its entries elementary symmetric polynomials (see e.g. \cite{Don} for
properties) depending on different sets of variables. We use the notation $%
\sigma ^{\pm }$ when they depend on the variable $x=\exp \theta $ associated
to the sets $M^{\pm }$ and $\hat{\sigma}$ to indicate that all variables are
multiplied by a factor $ie^{-\sigma }$. The overall constant was computed to 
\begin{eqnarray}
H_{2s+\tau ,2t+\tau ^{\prime }}^{{\cal O}_{\tau ,\tau ^{\prime }}^{\mu ,\nu
}|M^{+}M^{-}} &=&i^{s(2\tau +\tau ^{\prime }+\nu +2)}2^{s(2s-2t-\tau
^{\prime }-1+2\tau )}  \nonumber \\
&&\times e^{s\sigma (2t+\tau ^{\prime })/2}H_{\tau ,2t+\tau ^{\prime }}^{%
{\cal O}_{\tau ,\tau ^{\prime }}^{\mu ,\nu }}\,,
\end{eqnarray}
where the value of $H_{\tau ,2t+\tau ^{\prime }}^{{\cal O}_{\tau ,\tau
^{\prime }}^{\mu ,\nu }}$ is fixed by the lowest non-vanishing form factor.
In particular we need 
\begin{equation}
F_{2s,2t}^{\Theta }=\sigma _{1}(x_{1},\ldots ,x_{n})\sigma
_{1}(x_{1}^{-1},\ldots ,x_{n}^{-1})F_{2s,2t}^{{\cal O}_{2,2}^{1,1}}\,\,.
\end{equation}
Having assembled all the ingredients we can evaluate the expressions (\ref
{cr0}) and (\ref{dr0}). We carry out the integrals by means of a Monte Carlo
computation. For $c(r_{0})$ we take contributions up to the 4-particle form
factor into account and display our results in figure 1.

\begin{center}
\includegraphics[width=8.2cm,height=6.09cm]{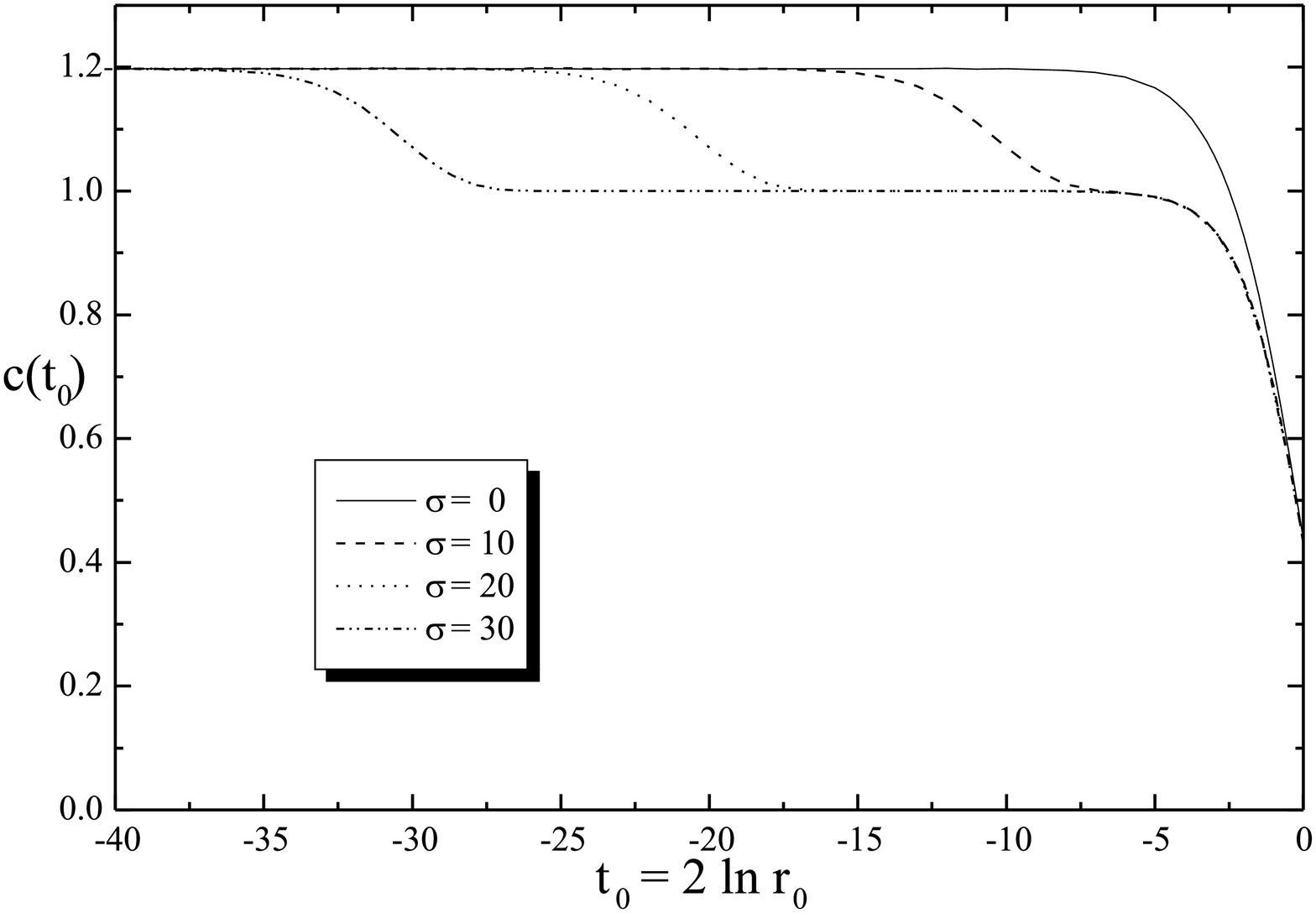}
\end{center}

\noindent {\small Figure 1: Renormalization group flow for the Virasoro
central charge $c(r_{0})$ for various values of the resonance parameter $%
\sigma$. }

\vspace*{1.2mm}Following the renormalization group flow from the ultraviolet
to the infrared, figure 1 illustrates the flow from the $SU(3)_{2}/U(1)^{2}$%
- to the $SU(2)_{2}/U(1)\otimes SU(2)_{2}/U(1)$-coset when the unstable
particle becomes massive. This confirms qualitatively the previous
observation of the TBA analysis \cite{CFKM}. Here we also want to compare
the value of the mass of the unstable particle at different points of the
resonance parameter $\sigma $ and $t_{0}$. \ Taking now the mass scales of
the stable particles to be the same, i.e. $m_{+}=m_{-}=m$, we compute the
mass of the unstable particle according to (\ref{Munst}), i.e. $M_{\tilde{u}%
}(t_{u},\sigma )\sim m/\sqrt{2}\exp ((\left| \sigma \right| +t_{u})/2)$.
This means for different values of the resonance parameter we may still have
the same value for the mass of the unstable particle when changing $t_{u}$,
indeed we find 
\begin{equation}
M_{\tilde{u}}(-30.8,30)=M_{\tilde{u}}(-20.8,20)=M_{\tilde{u}}(-10.8,10).
\label{umass}
\end{equation}
Since the flow between the two cosets is smooth and takes place over some
range of $t_{0}$, we had to select one particular point $t_{u}$. As already
indicated in general, it is convenient to identify $M_{\tilde{u}}$ as the
point at which $c(t_{0})$ is half the difference between the two coset
values of $c$. It is clear from figure 1, that since the overall shape of
the curves between two values of $c$ is identical for different values of $%
\sigma $, any other value in the interval would lead to the same results in
comparative considerations. This also means that when evaluating (\ref{umass}%
) the resulting value $0.47m$, which apparently violates the energetically
necessary condition $M_{\tilde{u}}>m_{a}+m_{b}$, should not be taken too
literally since the point $t_{u}$ is only chosen because it is easy to fix.
Equations (\ref{umass}) confirm our general assertions outlined above.

For the evaluation of the scaled conformal dimension (\ref{dr0}) we proceed
similarly. For the solutions corresponding to the operators ${\cal O}%
_{0,0}^{0,0}$, ${\cal O}_{0,2}^{0,1}${\small \ }and ${\cal O}_{2,0}^{1,0}$,
whose conformal dimension in the ultraviolet limit was identified \cite{CF}
to be $1/10$, we take up to the 6-particle form factors into account. For
the former two operators our results are presented in figures 2 and 3.

\begin{center}
\includegraphics[width=8.2cm,height=6.09cm]{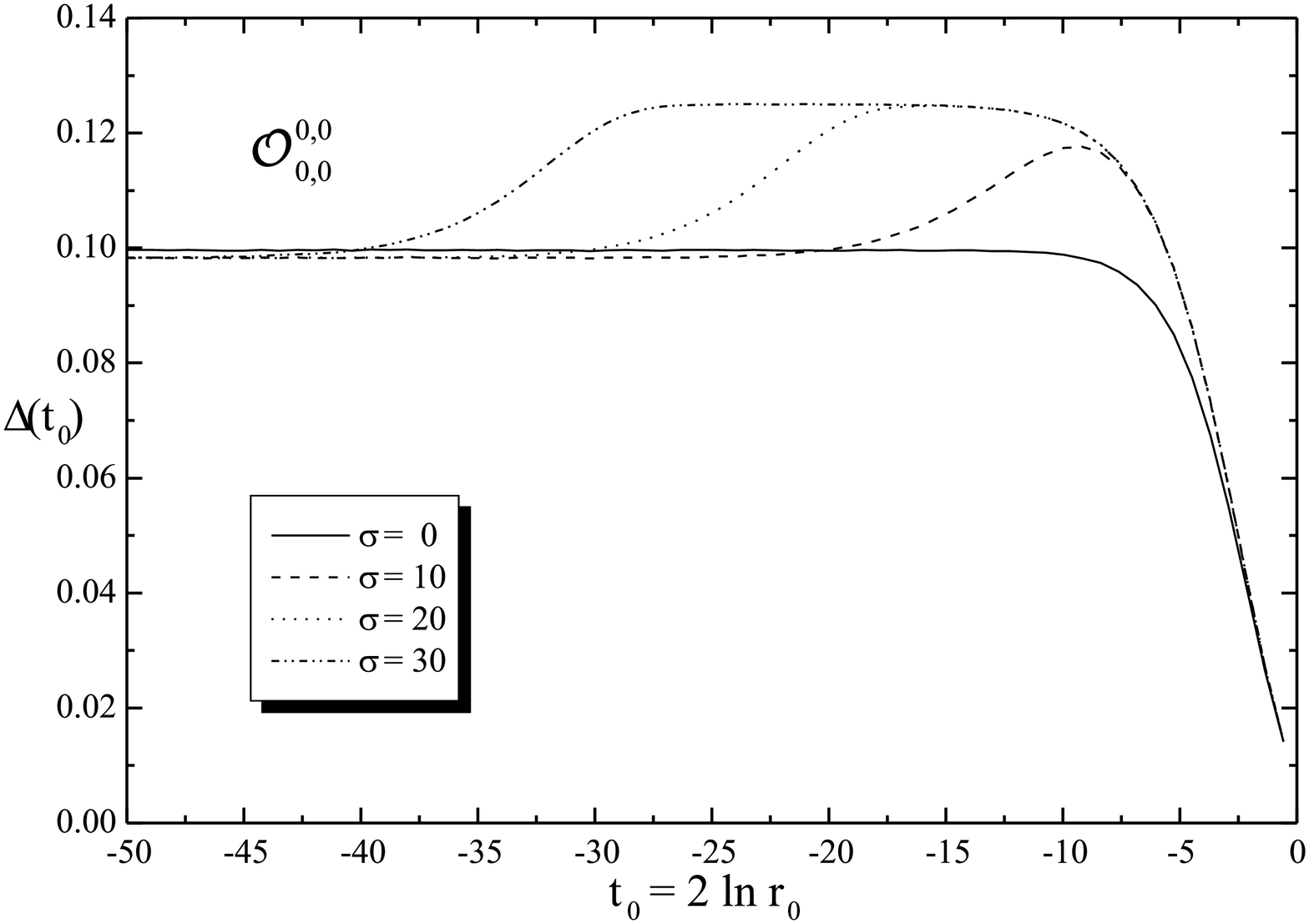}
\end{center}

\vspace*{0.2cm}

\noindent {\small Figure 2: Renormalization group flow for the conformal
dimension $\Delta (r_{0})$ of the operator ${\cal O}_{0,0}^{0,0} $ for
various values of the resonance parameter $\sigma$.}

\vspace*{1.2mm}We observe that the conformal dimension of the operator $%
{\cal O}_{0,0}^{0,0}$ flows to the value $1/8$, which is twice the conformal
dimension of the disorder operator $\mu $ in the Ising model. The factor $2$
is expected from the mentioned coset structure, i.e. we find two copies of $%
SU(2)_{2}/U(1)$. The nature of the operator is also anticipated, since by
construction $F_{n}^{{\cal O}_{0,0}^{0,0}|M^{+}M^{-}}$ of the $SU(3)_{2}$%
-HSG model coincides precisely with $F_{n}^{\mu }$ of the thermally
perturbed Ising model when one of the sets $M^{\pm }$ is empty. It is also
clear that we could alternatively obtain (\ref{umass}) from the analysis of $%
\Delta (r_{0})$.

\begin{center}
\includegraphics[width=8.2cm,height=6.09cm]{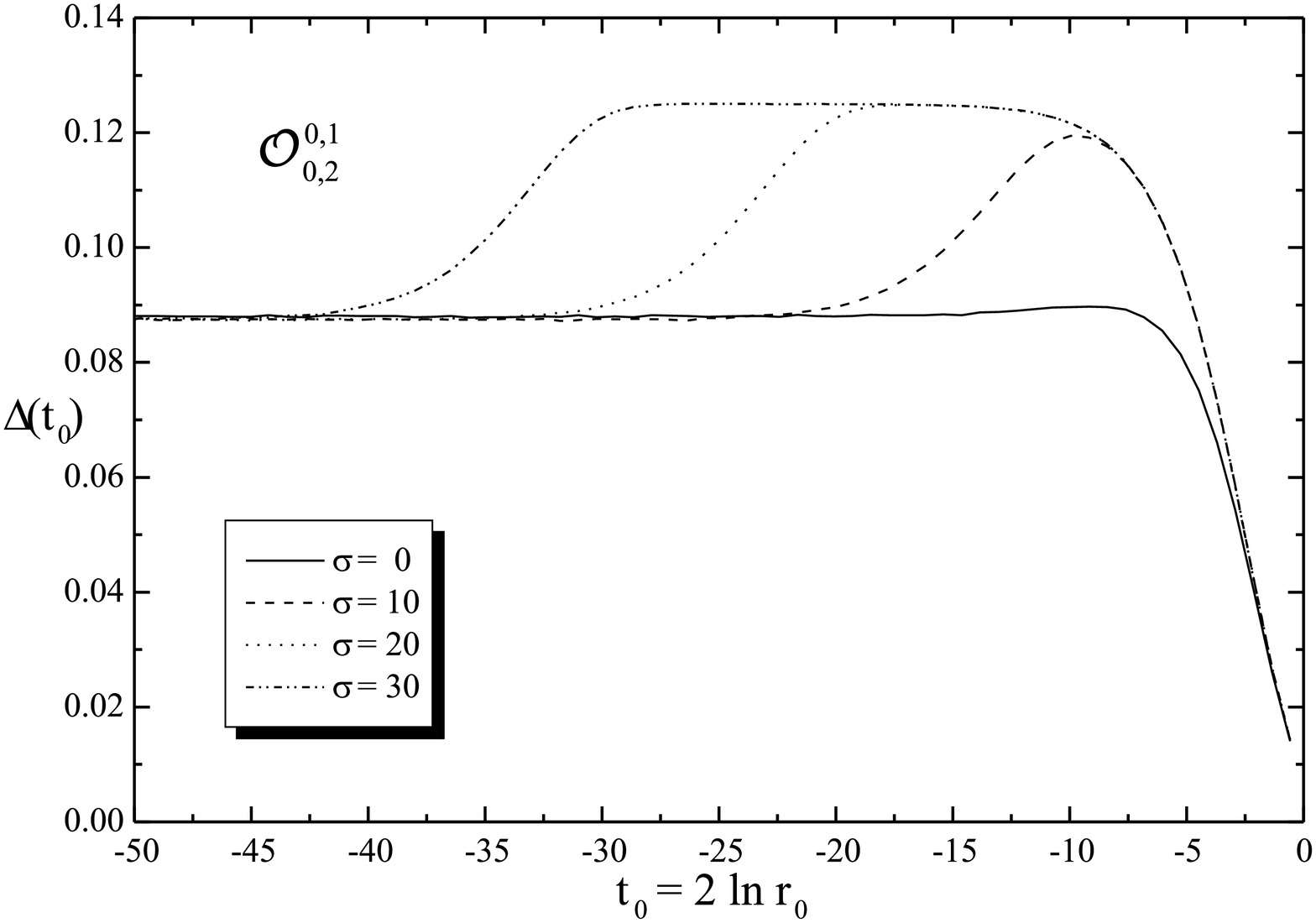}
\end{center}

\vspace*{0.2cm}

\noindent {\small Figure 3: Renormalization group flow for the conformal
dimension $\Delta (r_{0})$ of the operator ${\cal O}_{0,2}^{0,1} $ for
various values of the resonance parameter $\sigma$.}

\vspace*{1.2mm}

Despite the fact that the explicit expressions for \ the form factors of $%
{\cal O}_{0,2}^{0,1}${\small \ }and ${\cal O}_{2,0}^{1,0}${\small \ }differ
the values of $\Delta (r_{0})$ are hardly distinguishable and we therefore
omit the plots for the latter case. We also note the previously observed
fact \cite{CF}, that the higher particle contributions for the latter
operators are more important than for ${\cal O}_{0,0}^{0,0}$, which explains
the fact that the starting point at the ultraviolet fixed point is not quite 
$0.1$. The operators also flow to the value $1/8$, such that the degeneracy
of the $SU(3)_{2}$-HSG model disappears surjectively when the unstable
particles become massive.

In comparison with other methods it would be extremely desirable to
elaborate on the precise relationship between $c(r_{0})$ and the finite size
scaling function of the thermodynamic Bethe ansatz. Also the relation to the
intriguing proposal in \cite{Foda} of a renormalization group flow between
Virasoro characters remains unclarified. The analogue of $\Delta (r_{0})$
still needs to be identified in the TBA as well as in the context of \cite
{Foda}. In addition one may pose the question whether there exist higher
dimensional counterparts of the function $\Delta (r_{0})$ in analogy to the
results obtained in \cite{C4D} for $c(r_{0})$. Concerning the specific
status of the HSG-models it remains a challenge to extend the results to
other Lie groups \cite{CF2}.

\smallskip

\noindent {\bf Acknowledgments: } A.F. is grateful to the Deutsche
Forschungsgemeinschaft (Sfb288) for financial support. O.A.C. thanks CICYT
(AEN99-0589), DGICYT (PB96-0960), and the EC Commission (TMR grant
FMRX-CT96-0012) for partial financial support and is also very grateful to
the Institut f\"{u}r theoretische Physik of the Freie Universit\"{a}t for
hospitality and for partial financial support. We are grateful to \ J.L.
Miramontes, G. Mussardo for useful comments and A. Schilling for discussions
on \cite{Foda}.

\end{document}